\documentclass[a4paper,conference]{IEEEtran}
\usepackage{cite}

\usepackage{graphicx}
\usepackage[cmex10]{amsmath}
\usepackage{amssymb}
\usepackage{wasysym}

\usepackage{booktabs}
\usepackage[tight,footnotesize]{subfigure}

\usepackage{url}

\usepackage{relsize}
\usepackage[export]{adjustbox}
\usepackage{lipsum} 
\usepackage{color}

\usepackage{CJKutf8}
\usepackage{bm}
\usepackage[binary-units=true]{siunitx}

\begin{document}
%
\title{A High-Resolution, Wave and Current Resource Assessment of Japan: The Web GIS Dataset}

\author{\IEEEauthorblockN{Adrean Webb\IEEEauthorrefmark{1},
Takuji Waseda\IEEEauthorrefmark{1},
Wataru Fujimoto\IEEEauthorrefmark{1},
Kazutoshi Horiuchi\IEEEauthorrefmark{3},
Keiji Kiyomatsu\IEEEauthorrefmark{3},\\
Kazuhiro Matsuda\IEEEauthorrefmark{4},
Yasumasa Miyazawa\IEEEauthorrefmark{2},
Sergey Varlamov\IEEEauthorrefmark{2},
and
Jun Yoshikawa\IEEEauthorrefmark{4}
}
\vspace{1.5ex}
\IEEEauthorblockA{\IEEEauthorrefmark{1}\textit{Department of Ocean Technology, Policy, and Environment}\\
\textit{The University of Tokyo, 5-1-5 Kashiwanoha, Kashiwa, Chiba, Japan}\\ 
{\tt\small webb@isea.k.u-tokyo.ac.jp, waseda@k.u-tokyo.ac.jp, fujimoto@isea.k.u-tokyo.ac.jp}}
\vspace{1.ex}
\IEEEauthorblockA{\IEEEauthorrefmark{2}\textit{JAMSTEC-APL, 
3173-25 Showa-machi, Kanazawa-ku, Yokohama-city, Kanagawa, Japan}\\
{\tt\small miyazawa@jamstec.go.jp, vsm@jamstec.go.jp}}
\vspace{1.ex}
\IEEEauthorblockA{\IEEEauthorrefmark{4}\textit{Web-Brain, 2-3-6 Sotokanda, Chiyoda-ku, Tokyo, Japan}\\
{\tt\small kzmat@web-brain.jp, jun@web-brain.jp}}
\vspace{1.ex}
\IEEEauthorblockA{\IEEEauthorrefmark{3}\textit{Forecast Ocean Plus, 2-2-5 Kudanminami, Chiyoda-ku, Tokyo, Japan}\\
{\tt\small horiuchi@forecastocean.com, kiyomatsu@forecastocean.com}}
}

\maketitle

\begin{abstract}
%
The University of Tokyo and JAMSTEC have conducted state-of-the-art wave and current resource assessments to assist with generator site identification and construction in Japan.
%
%
These assessments are publicly-available and accessible via a web GIS service designed by WebBrain that utilizes TDS and GeoServer software with Leaflet libraries.
%
%
%
The web GIS dataset contains statistical analyses of wave power, ocean and tidal current power, ocean temperature power, and other basic physical variables.
%
%
The data (2D maps, time charts, depth profiles, etc.) is accessed through interactive browser sessions and downloadable files.
\end{abstract}

\begin{IEEEkeywords}
wave resource assessment, ocean and tidal current resource assessment, ocean temperature resource assessment, web GIS, 
NOAA WAVEWATCH~III, \mbox{JCOPE-T}
\end{IEEEkeywords}

%
\IEEEpeerreviewmaketitle

\section{Introduction}
With an extensive coastline and close proximity of the Kuroshio current, Japan is an ideal candidate for wave and ocean current energy extraction. 
Supported by New Energy and Industrial Technology Development Organization (NEDO), The University of Tokyo and JAMSTEC have conducted state-of-the-art wave and current resource assessments to assist with generator site identification and construction in Japan. 
%
These marine renewable energy assessments are now publicly-available and accessible via a web GIS service designed by WebBrain and maintained by The University of Tokyo.

An English interface of the web GIS dataset is located at \url{http://www.todaiww3.k.u-tokyo.ac.jp/nedo_p/en}. 
The resource assessments are located within the `webgis' subdirectory, which can be accessed via the `Marine Energy Web GIS' menu button on the left side (\url{http://www.todaiww3.k.u-tokyo.ac.jp/nedo_p/en/webgis/}).
Useful information and references can also be found within the menu button categories `Manual', `FAQ', and `Papers'.
%

\section{Resource assessments}
For full details of the resource assessments, please see \cite{WasedaWebb2016} and \cite{WebbWaseda2016}, and other references listed under the `Papers' category (\url{http://www.todaiww3.k.u-tokyo.ac.jp/nedo_p/en/papers/}).
To facilitate easy reference, relevant variables and definitions from the assessments are tabulated in Table~\ref{tab:Terminology}.

\begin{table}[!t]
\renewcommand{\arraystretch}{1.3}
\caption{List of relevant variables and definitions.}
\label{tab:Terminology}
\centering\small
\begin{tabular}{p{0.21\columnwidth}p{0.1\columnwidth}p{0.55\columnwidth}}
\toprule
\bf Terminology & \bf Unit & \bf Long name \\
\midrule
$h_{m0}$ & \si{\meter} & Spectral significant wave height \\
$f_p$ & \si{\hertz} & Peak frequency \\
$t_{-1}$ & \si{\second} & Wave energy period (inverse moment) \\
$\theta_w$ & \si{deg} & Mean wave direction (meteorological)\\
$\bm U_{10}$, $U_{10}$, $V_{10}$ & \si{\meter \per \second} & \SI{10}{\meter} wind vector and components \\
$P_{w1}$ & \si{\kilo \watt \per \meter} & Deep-water wave power density \\
$P_{w3}$ & \si{\kilo \watt \per \meter} & Doppler-shifted wave power density \\
\midrule
$\bm u$, $u$, $v$ & \si{\meter \per \second} & Current vector and components \\
$\theta_c$ & \si{deg} & Current direction (oceanographic)\\
$m_c$ & \si{\percent} & Current flow direction stability \\
$P_c$ & \si{\watt \per \meter^2} & Ocean current power density \\
\midrule
$T$ & \si{\celsius} & Water temperature \\
$\Delta T_{20}$ & \si{\celsius} & Water temperature difference with \SI{20}{\meter} depth\\
$P_T$ & \si{\watt \per \meter^2} & Ocean thermal power density \\
\midrule
$J\left[X,Y\right]$ & \si{\percent} & Discrete joint probability distribution of $X$ and $Y$\\
$F\left[X\right]$ & \si{\percent} & Discrete frequency distribution of $X$\\
$F_c\left[X \geq x \right]$ & \si{\percent} & Relative cumulative frequency of $X$ greater than or equal to $x$\\
\bottomrule
\end{tabular}
\end{table}
\begin{figure*}[!t]
\centering
\subfigure[Interactive map of NOWPHAS observation locations]{
\includegraphics[width=0.88\columnwidth,frame]{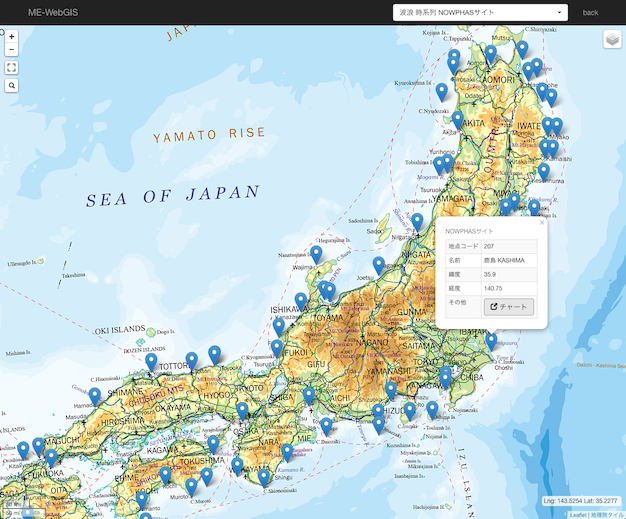}}
\hspace{0.05\columnwidth}
\subfigure[Interactive map of a test site (Kamaishi)]
{\includegraphics[width=0.88\columnwidth,frame]{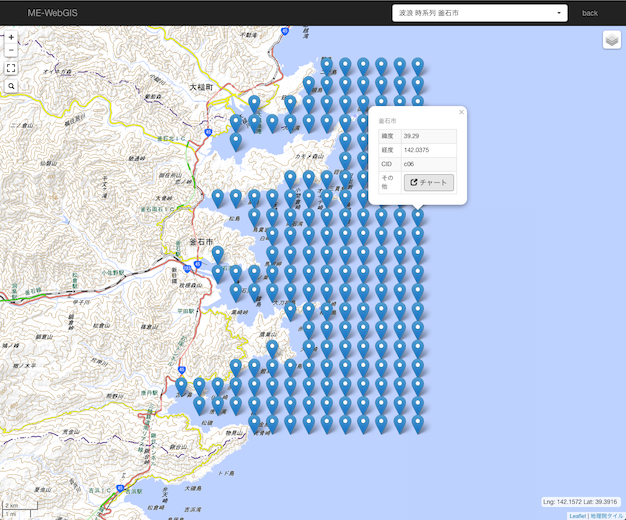}}
\subfigure[{Sample download of $J \left[ h_{m0}, t_{-1} \right]$}]
{\includegraphics[width=0.88\columnwidth,frame]{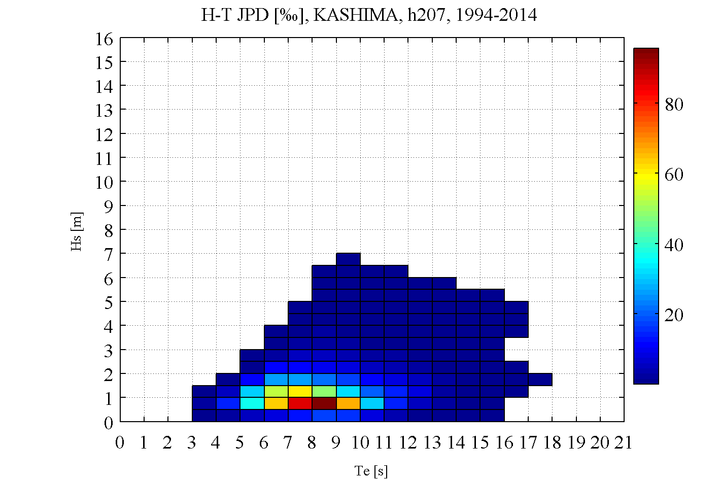}}
\hspace{0.05\columnwidth}
\subfigure[{Sample download of an interactive wave time series ($P_{w3}$)}]{
\includegraphics[width=0.88\columnwidth,frame]{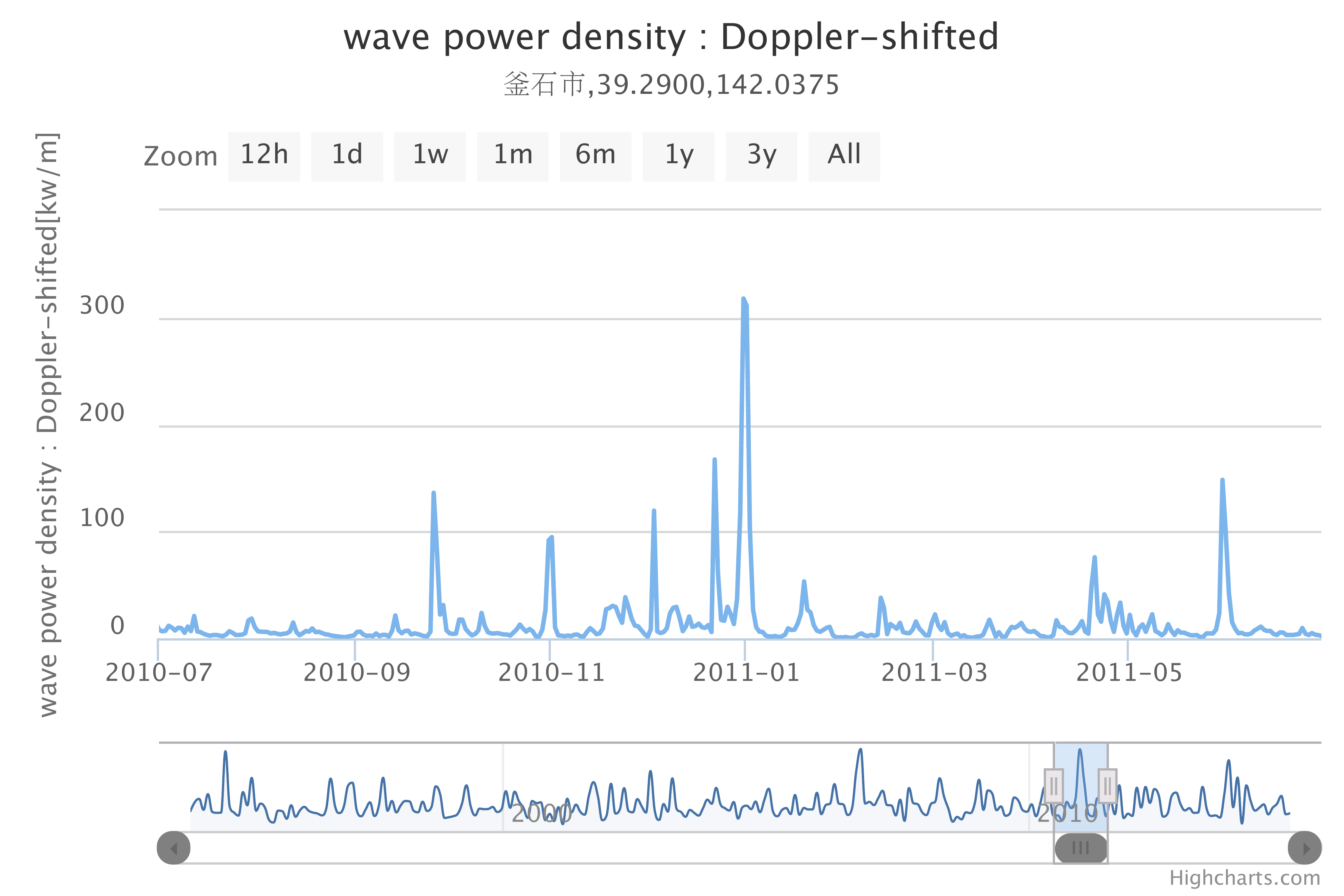}}
\caption{
Sample of TodaiWW3 time series data available for 21~years (1994--2014). 
Screenshots of interactive maps are shown in (a) and (b) for the 82 NOWPHAS locations and 11 test sites (respectively).
%
%
Sample file downloads (for the selected sites) are shown in (c) joint probability distribution of significant wave height and wave energy period and (d) interactive time series of $P_{w3}$.
}
\label{fig:WaveTimeSeries}
\end{figure*}
\subsection{Wave resource assessment}
%
The wave resource assessment (University of Tokyo) is based on a large-scale NOAA WAVEWATCH III (version~4.18) simulation that includes several key components to improve model skill, such as a four-nested-layer approach for high resolution and the addition of current effects in all wave power density calculations. 
%
The nearshore resolution of each model grid (24 on the lowest nest) is approximately \SI{1}{\kilo\meter}, with a latitude-longitude grid cell size of $\ang{0.010} \times \ang{0.0125}$. 
Data from all the grids are post-processed and assembled into a single grid, referred to as \emph{TodaiWW3} here.
%
The main simulation covers a 21-year period (1994/01/0 to 2014/12/31) to reduce aleatory uncertainty \cite{KidouraWada2014} and required approximately 3~million parallel process hours to run on The University of Tokyo FX10 supercomputer.

\subsection{Current and temperature resource assessments}
The ocean and tidal current and ocean temperature resource assessments (JAMSTEC) are based on a high-resolution simulation using JCOPE-T -- a regional, $1/36^{\circ}$ tide-resolving model with 47 sigma levels (approximately \SI{3}{\kilo\meter} in the horizontal) that is nested within the data-assimilating, $1/12^{\circ}$ non-tidal \mbox{JCOPE2} model. 
The simulation covers a \mbox{10-year} period (2002/01/01 to 2011/12/31) and 
required approximately 25~thousand parallel process hours to run on a JAMSTEC NEC SX-9 supercomputer.
%
%
Both assessments are referred to as \mbox{\emph{JCOPE-T}} here.
\section{Data distribution}
%

%

The marine renewable energy resource assessments are distributed using web GIS, a newly-developed web mapping service.
%
%
The web GIS engine is a THREDDS Data Server (TDS) that provides remote access to the datasets (http://doi.org/10.5065/D6N014KG).
Topographical data is handled using GeoServer, an open geospatial information server (http://geoserver.org), and 
the user interface utilizes `Leaflet', a browser and mobile mapping library (http://leafletjs.com/).
%
The server comprises a 12-Core \SI{2.5}{\giga \hertz} CPU (Xeon E5-2680v3), \SI{64}{\giga \byte} RAM, \SI{800}{\giga \byte} SSD, and $5{\times}4$\si{\tera \byte} HD (RAID5) hardware unit running CentOS 6.6; it is projected to handle a 10 concurrent user load.

The web GIS dataset is accessed through interactive browser sessions and downloadable files, and consists primarily of NetCDF files (COARDS convention).
%
These files contain the mean and climatology data (and other related variables) that are used to create the interactive 2D maps. 
%
%
Location-specific analyses are also available, such as time series, joint probability distributions, time charts, and depth profiles.
%

\begin{figure}[t!]
\centering
%
\subfigure[Monthly mean of $h_{m0}$ for 2014/12 (no threshold)]
{\includegraphics[width=0.88\columnwidth,frame]{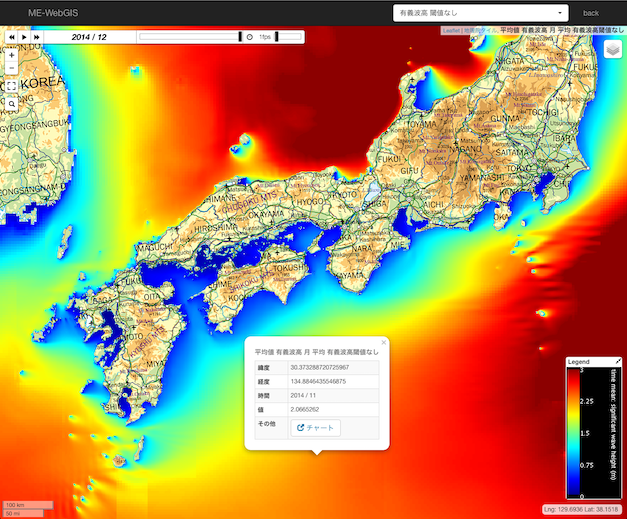}}
\subfigure[Monthly standard deviation of $h_{m0}$ for 2014/12 (no threshold)]
{\includegraphics[width=0.88\columnwidth,frame]{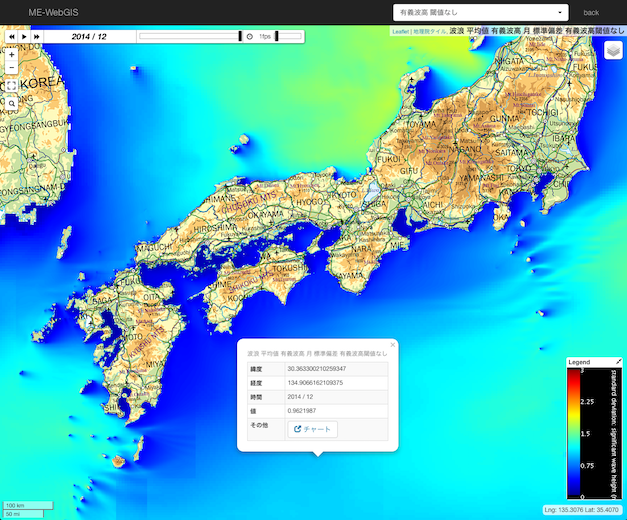}}
\hspace{0.05\columnwidth}
\subfigure[Time chart of $h_{m0}$ monthly means and standard deviations]
{\includegraphics[width=0.88\columnwidth,frame]{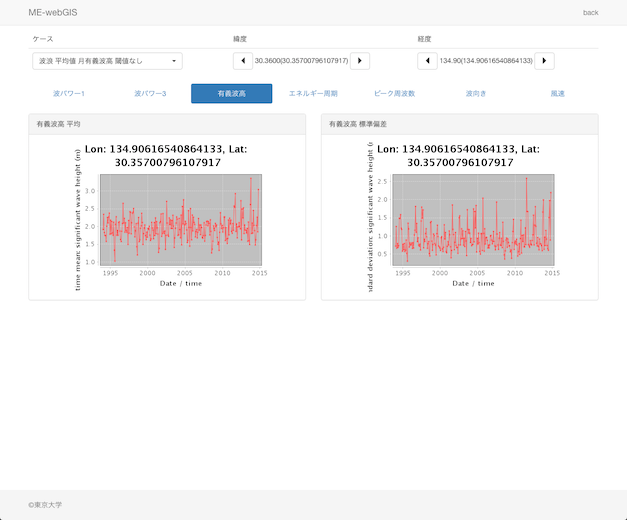}}
\caption{
Sample of TodaiWW3 mean data available for each of the 21~years (1994--2014).
Screenshots of interactive monthly outputs of $h_{m0}$ are shown for (a) mean and (b) standard deviation.
(c) Means and standard deviations can be viewed as time charts at each map location.
}
\label{fig:WaveMapTime}
\end{figure}

\begin{figure}[!t]
\centering
\subfigure[Winter climatology of $P_{w3}$ with no $h_{m0}$ threshold]
{\includegraphics[width=0.88\columnwidth,frame]{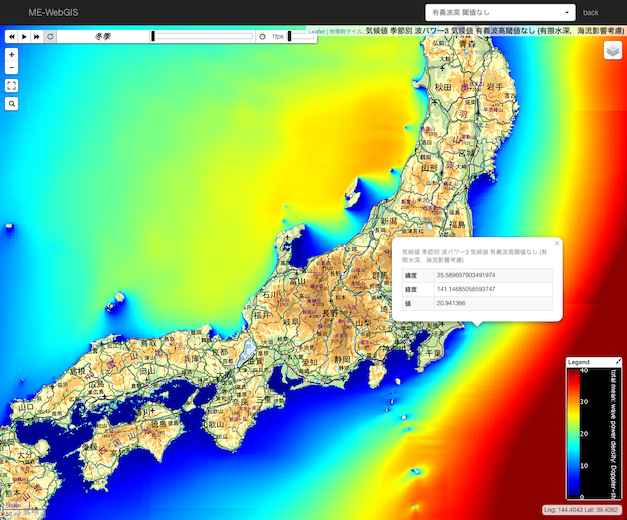}}
\hspace{0.05\columnwidth}
\subfigure[Winter climatology of $P_{w3}$ with a \SI{5}{\meter} $h_{m0}$ threshold]
{\includegraphics[width=0.88\columnwidth,frame]{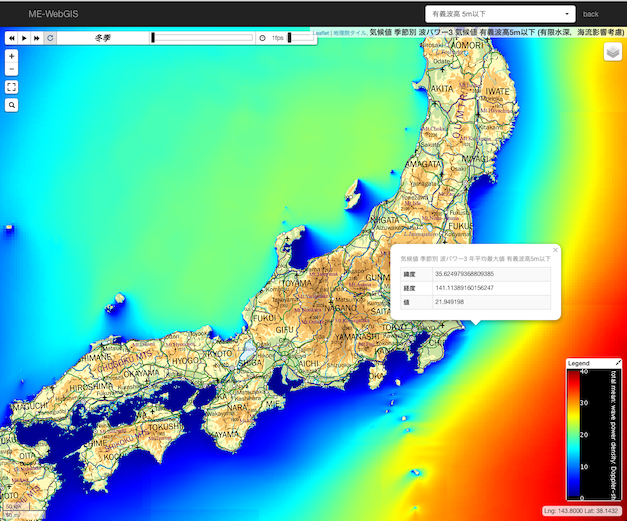}}
%
\subfigure[Winter mean minimum of $P_{w3}$ with a \SI{5}{\meter} $h_{m0}$ threshold]
{\includegraphics[width=0.88\columnwidth,frame]{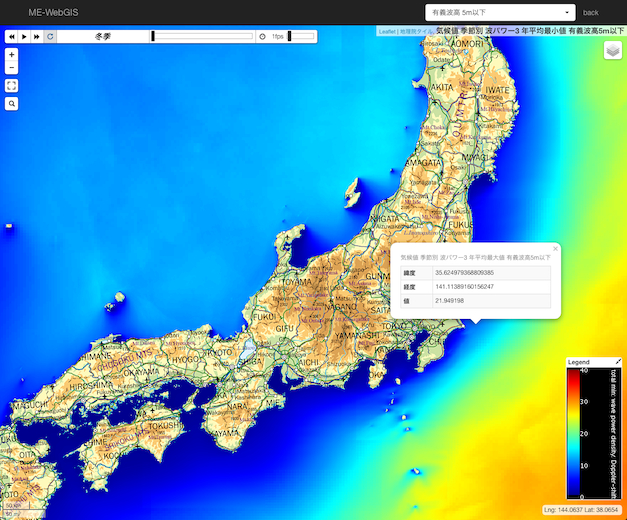}}
\caption{
Sample of TodaiWW3 climatology map data available for a 21-year period (1994--2014).
%
Screenshots of interactive seasonal outputs of $P_{w3}$ are shown for two thresholds for (a) climatology (none), (b) climatology (\SI{5}{\meter}), and (c) mean maximum (\SI{5}{\meter}).
%
%
%
}
\label{fig:WaveMapClim}
\end{figure}

Within each interactive 2D map session, drop-down menus in the top right corner are used to access or change the relevant analyses (mean period type, depth, etc.).
%
A layer icon in the same corner is used to select the desired variable and background geographical map.
%
%
For some outputs, a time display navigator in the top left corner is used to select the desired time period and can be used to automatically cycle through all periods.
%
%
Clicking on any location will reveal in a pop-up display the coordinates, time (if relevant), and value (`none' if on land or outside the domain).
Location-specific analyses can be accessed via the link in these displays.
A search bar in the top left corner can also be used to search for locations by name. 
See Figs.~\ref{fig:WaveMapTime}--\ref{fig:OceanMap} for sample screenshots of interactive sessions.
\section{Wave time series information}
\begin{table}[!t]
\renewcommand{\arraystretch}{1.3}
\caption{
Hourly TodaiWW3 time series data is available for 82 NOWPHAS locations and 11 NEDO test sites for 21~years (1994--2014).
%
%
Duration of the time series is adjustable and downloadable.
%
In addition, the joint probability distribution of significant wave height and wave energy period is provided and downloadable.
%
%
}
\label{tab:WaveTimeSeries}
\centering\small
\begin{tabular}{p{0.18\columnwidth}p{0.1\columnwidth}p{0.1\columnwidth}p{0.1\columnwidth}p{0.1\columnwidth}}
\toprule
\bf Variable & \multicolumn{2}{c}{\bf NOWPHAS} & \multicolumn{2}{c}{\bf Test sites} \\
& \multicolumn{1}{c}{Histogram} & \multicolumn{1}{c}{Series} & \multicolumn{1}{c}{Histogram} & \multicolumn{1}{c}{Series}\\
\midrule
$h_{m0}$ & & \multicolumn{1}{c}{$\circ$} & & \multicolumn{1}{c}{$\circ$} \\
$f_p$ & & \multicolumn{1}{c}{$\circ$} & & \multicolumn{1}{c}{$\circ$} \\
$t_{-1}$ & & \multicolumn{1}{c}{$\circ$} & & \multicolumn{1}{c}{$\circ$} \\
$\theta_w$ & & \multicolumn{1}{c}{$\circ$} & & \multicolumn{1}{c}{$\circ$} \\
$P_{w1}$ & & \multicolumn{1}{c}{$\circ$} & & \multicolumn{1}{c}{$\circ$} \\
$P_{w3}$ & & \multicolumn{1}{c}{$\circ$} & & \multicolumn{1}{c}{$\circ$} \\
$J \left[h_{m0},t_{-1} \right]$ & \multicolumn{1}{c}{$\circ$} & & \multicolumn{1}{c}{$\circ$} & \\
\bottomrule
\end{tabular}
\end{table}
%

Access to the TodaiWW3 time series data is located within the subcategory `Wave Time Series Information' (\url{http://www.todaiww3.k.u-tokyo.ac.jp/nedo_p/jp/webgis/#i}).
%
Within it, hourly time series data is available for 82 observational locations and 11 NEDO test sites.
The observational locations chosen coincide with a coastal observational network maintained by Nationwide Ocean Wave Information for Ports and HArborS (NOWPHAS).
Observational data is distributed by NOWPHAS (\url{http://nowphas.mlit.go.jp/}) and is not included in the web GIS dataset.
Respectively, each NEDO test site covers a $\ang{0.2} \times \ang{0.2}$ latitude-longitude region (approximately 300 grid cell locations).

Observational and test site locations are chosen via interactive browser sessions such as the sample screenshots shown in Figs.~\ref{fig:WaveTimeSeries}a--b.
Each selected location contains six time series with adjustable durations in UTC format.
See Table~\ref{tab:WaveTimeSeries} for a full list of variables and Fig.~\ref{fig:WaveTimeSeries}d for a sample download.
The 21-year joint probability distributions of significant wave height and the wave energy period are also provided; see Fig.~\ref{fig:WaveTimeSeries}c for a sample download.
 %
%
The time series data can be downloaded as a `png', `jpeg', `pdf', `svg', and `csv' files while the joint probability distributions can be downloaded as `png' and `csv' files.
\section{Wave map information}\label{sec:WaveMap}
Access to the TodaiWW3 mean and climatology data is located within the subcategory `Wave Map Information' (\url{http://www.todaiww3.k.u-tokyo.ac.jp/nedo_p/jp/webgis/#i-2}) and is further subdivided by means and climatologies.
Nine variables are chosen and analyzed for online distribution; see Table~\ref{tab:WaveList} for a full list.
Since there is a high variability in wave power density near Japan, five significant wave height thresholds -- \SI{3}{\meter}, \SI{5}{\meter}, \SI{7}{\meter}, \SI{9}{\meter}, and none -- are applied when calculating each statistic and any corresponding value above this limit is ignored (see Table~\ref{tab:WaveList} for applicability).
As an example, winter climatologies of the Doppler-shifted wave power density are shown both with a \SI{5}{\meter} threshold and without in Figs.~\ref{fig:WaveMapClim}a--b.
\begin{table}[!t]
\renewcommand{\arraystretch}{1.2}
\caption{
2D maps and time charts (at every map location) of TodaiWW3 data are calculated for select period types and significant wave height thresholds.
%
Different mean/climatology types (year/season/month) are calculated for each/all of the 21~years (1994--2014).
%
%
%
Mean/climatology of $\bm U_{10}$ refers to $\left( \langle U_{10} \rangle, \langle V_{10} \rangle \right)$.
See Section~\ref{sec:WaveMap} for details.
}
\label{tab:WaveList}
\centering\small
\begin{tabular}{p{0.1\columnwidth}
p{0.02\columnwidth}p{0.02\columnwidth}p{0.02\columnwidth}
p{0.02\columnwidth}p{0.02\columnwidth}p{0.02\columnwidth}p{0.02\columnwidth}p{0.02\columnwidth}}
\toprule
\bf Variable 
& \multicolumn{8}{c}{\bf Mean / Climatology}\\
%
& \multicolumn{3}{c}{\textit{Type}}
& \multicolumn{5}{c}{\textit{Threshold ($h_{m0}$; \si{\meter})}} \\
& 
\multicolumn{1}{c}{Year} & \multicolumn{1}{c}{Season} & \multicolumn{1}{c}{Month} &
\multicolumn{1}{c}{None} & \multicolumn{1}{c}{$3$} & \multicolumn{1}{c}{$5$} & \multicolumn{1}{c}{$7$} & \multicolumn{1}{c}{$9$} \\
\midrule
$h_{m0}$ 
& \multicolumn{1}{c}{$\circ$} & \multicolumn{1}{c}{$\circ$} & \multicolumn{1}{c}{$\circ$} 
& \multicolumn{1}{c}{$\circ$} & \multicolumn{1}{c}{$\circ$} & \multicolumn{1}{c}{$\circ$} & \multicolumn{1}{c}{$\circ$} & \multicolumn{1}{c}{$\circ$} \\
$f_p$ 
& \multicolumn{1}{c}{$\circ$} & \multicolumn{1}{c}{$\circ$} & \multicolumn{1}{c}{$\circ$} 
& \multicolumn{1}{c}{$\circ$} & \multicolumn{1}{c}{$\circ$} & \multicolumn{1}{c}{$\circ$} & \multicolumn{1}{c}{$\circ$} & \multicolumn{1}{c}{$\circ$} \\
$t_{-1}$ 
& \multicolumn{1}{c}{$\circ$} & \multicolumn{1}{c}{$\circ$} & \multicolumn{1}{c}{$\circ$} 
& \multicolumn{1}{c}{$\circ$} & \multicolumn{1}{c}{$\circ$} & \multicolumn{1}{c}{$\circ$} & \multicolumn{1}{c}{$\circ$} & \multicolumn{1}{c}{$\circ$} \\
$\theta_w$ 
& \multicolumn{1}{c}{$\circ$} & \multicolumn{1}{c}{$\circ$} & \multicolumn{1}{c}{$\circ$} 
& \multicolumn{1}{c}{$\circ$} & \multicolumn{1}{c}{$\circ$} & \multicolumn{1}{c}{$\circ$} & \multicolumn{1}{c}{$\circ$} & \multicolumn{1}{c}{$\circ$}  \\
$\bm U_{10}$
& \multicolumn{1}{c}{$\circ$} & \multicolumn{1}{c}{$\circ$} & \multicolumn{1}{c}{$\circ$} 
& \multicolumn{1}{c}{$\circ$} & & & & \\
$U_{10}$
& \multicolumn{1}{c}{$\circ$} & \multicolumn{1}{c}{$\circ$} & \multicolumn{1}{c}{$\circ$} 
& \multicolumn{1}{c}{$\circ$} & & & & \\
$V_{10}$
& \multicolumn{1}{c}{$\circ$} & \multicolumn{1}{c}{$\circ$} & \multicolumn{1}{c}{$\circ$} 
& \multicolumn{1}{c}{$\circ$} & & & & \\
$P_{w1}$ 
& \multicolumn{1}{c}{$\circ$} & \multicolumn{1}{c}{$\circ$} & \multicolumn{1}{c}{$\circ$} 
& \multicolumn{1}{c}{$\circ$} & \multicolumn{1}{c}{$\circ$} & \multicolumn{1}{c}{$\circ$} & \multicolumn{1}{c}{$\circ$} & \multicolumn{1}{c}{$\circ$} \\
$P_{w3}$ 
& \multicolumn{1}{c}{$\circ$} & \multicolumn{1}{c}{$\circ$} & \multicolumn{1}{c}{$\circ$} 
& \multicolumn{1}{c}{$\circ$} & \multicolumn{1}{c}{$\circ$} & \multicolumn{1}{c}{$\circ$} & \multicolumn{1}{c}{$\circ$} & \multicolumn{1}{c}{$\circ$} \\
\bottomrule
\end{tabular}
\end{table}
\begin{table}[!t]
\renewcommand{\arraystretch}{1.2}
\caption{
List of 2D maps and time charts (at every map location) available for TodaiWW3 mean and climatology data.
%
%
%
`Std' and `Clim' are shorthand for standard deviation and climatology;
`Max' and `Min' refer to the mean maximum and minimum at each location.
%
%
See Section~\ref{sec:WaveMap} and Table~\ref{tab:WaveList} for further details.
}
\label{tab:WaveTime}
\centering\small
\begin{tabular}{p{0.1\columnwidth}
p{0.02\columnwidth}p{0.02\columnwidth}
p{0.02\columnwidth}p{0.02\columnwidth}
p{0.02\columnwidth}p{0.02\columnwidth}p{0.02\columnwidth}
}
\toprule
\bf Variable 
& \multicolumn{4}{c}{\bf Mean} 
& \multicolumn{3}{|c}{\bf Climatology} 
\\
%
& \multicolumn{2}{c}{\textit{Map}} 
& \multicolumn{2}{c}{\textit{Chart}} 
& \multicolumn{3}{|c}{\textit{Map}} 
\\
& 
\multicolumn{1}{c}{Mean} & \multicolumn{1}{c}{Std} &
\multicolumn{1}{c}{Mean} & \multicolumn{1}{c}{Std} 
& \multicolumn{1}{|c}{Clim} & \multicolumn{1}{c}{Max} & \multicolumn{1}{c}{Min}
\\
\midrule
$h_{m0}$ 
& \multicolumn{1}{c}{$\circ$} & \multicolumn{1}{c}{$\circ$} & \multicolumn{1}{c}{$\circ$} & \multicolumn{1}{c}{$\circ$} 
& \multicolumn{1}{|c}{$\circ$} & \multicolumn{1}{c}{$\circ$} & \multicolumn{1}{c}{$\circ$} 
\\
$f_p$ 
& \multicolumn{1}{c}{$\circ$} & \multicolumn{1}{c}{$\circ$} & \multicolumn{1}{c}{$\circ$} & \multicolumn{1}{c}{$\circ$} 
& \multicolumn{1}{|c}{$\circ$} & \multicolumn{1}{c}{$\circ$} & \multicolumn{1}{c}{$\circ$} 
\\
$t_{-1}$ 
& \multicolumn{1}{c}{$\circ$} & \multicolumn{1}{c}{$\circ$} & \multicolumn{1}{c}{$\circ$} & \multicolumn{1}{c}{$\circ$} 
& \multicolumn{1}{|c}{$\circ$} & \multicolumn{1}{c}{$\circ$} & \multicolumn{1}{c}{$\circ$} 
\\
$\theta_w$ 
& \multicolumn{1}{c}{$\circ$} & \multicolumn{1}{c}{$\circ$} & \multicolumn{1}{c}{$\circ$} & \multicolumn{1}{c}{$\circ$} 
& \multicolumn{1}{|c}{$\circ$} & & 
\\
$\bm U_{10}$
& \multicolumn{1}{c}{$\circ$} & & \multicolumn{1}{c}{$\circ$} & 
& \multicolumn{1}{|c}{$\circ$} & & 
\\
$U_{10}$
& \multicolumn{1}{c}{$\circ$} & & \multicolumn{1}{c}{$\circ$} & 
& \multicolumn{1}{|c}{$\circ$} & & 
\\
$V_{10}$
& \multicolumn{1}{c}{$\circ$} & & \multicolumn{1}{c}{$\circ$} & 
& \multicolumn{1}{|c}{$\circ$} & & 
\\
$P_{w1}$ 
& \multicolumn{1}{c}{$\circ$} & \multicolumn{1}{c}{$\circ$} & \multicolumn{1}{c}{$\circ$} & \multicolumn{1}{c}{$\circ$} 
& \multicolumn{1}{|c}{$\circ$} & \multicolumn{1}{c}{$\circ$} & \multicolumn{1}{c}{$\circ$} 
\\
$P_{w3}$ 
& \multicolumn{1}{c}{$\circ$} & \multicolumn{1}{c}{$\circ$} & \multicolumn{1}{c}{$\circ$} & \multicolumn{1}{c}{$\circ$} 
& \multicolumn{1}{|c}{$\circ$} & \multicolumn{1}{c}{$\circ$} & \multicolumn{1}{c}{$\circ$} 
\\
\bottomrule
\end{tabular}
\end{table}
%

%
For each variable and threshold in Table~\ref{tab:WaveList}, three different types of means or climatologies -- annual, seasonal, and monthly (referred to as \emph{year}, \emph{season}, and \emph{month} hereafter) -- are calculated for each or all of the 21~years. 
Winter, spring, summer, and autumn seasons are classified as months December--February, March--May, June--August, and September--November respectively.
Means and standard deviations (for select variables) are available both as 2D maps and time charts (at map locations); see Table~\ref{tab:WaveTime} for a list and Figs.~\ref{fig:WaveMapTime}a--c for examples.
%
%
The 21-year climatologies are available as 2D maps.
In addition, mean maximums or minimums (at each map location) are available (for select variables) for their respective period type (year/season/month); see Table~\ref{tab:WaveTime} for a list and Fig.~\ref{fig:WaveMapClim}c for a mean maximum example.
\section{Current and temperature map information}\label{sec:OceanMap}
\begin{figure}[!t]
\centering
\subfigure[Spring climatology of $P_{c}$ at a \SI{1}{\meter} depth]
{\includegraphics[width=0.88\columnwidth,frame]{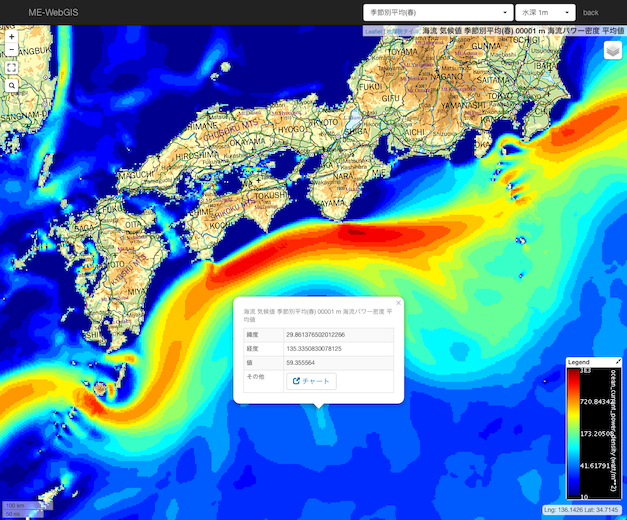}}
\hspace{0.05\columnwidth}
\subfigure[Large-meander climatology of $\bm u$ at a \SI{1}{\meter} depth]
{\includegraphics[width=0.88\columnwidth,frame]{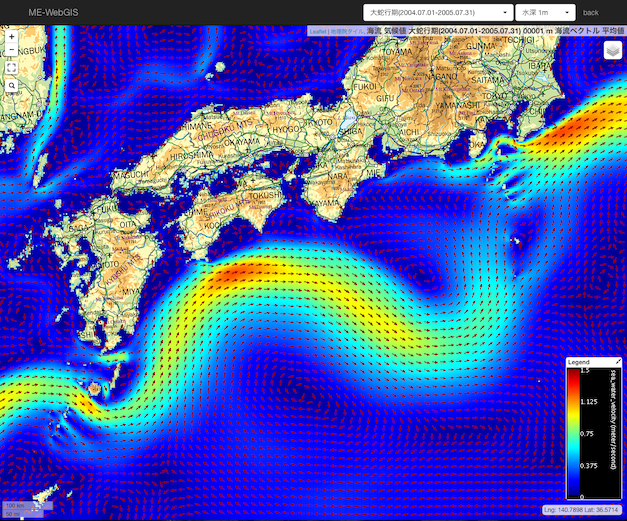}}
\hspace{0.05\columnwidth}
\subfigure[Annual climatology of $P_{T}$ at a \SI{200}{\meter} depth]
{\includegraphics[width=0.88\columnwidth,frame]{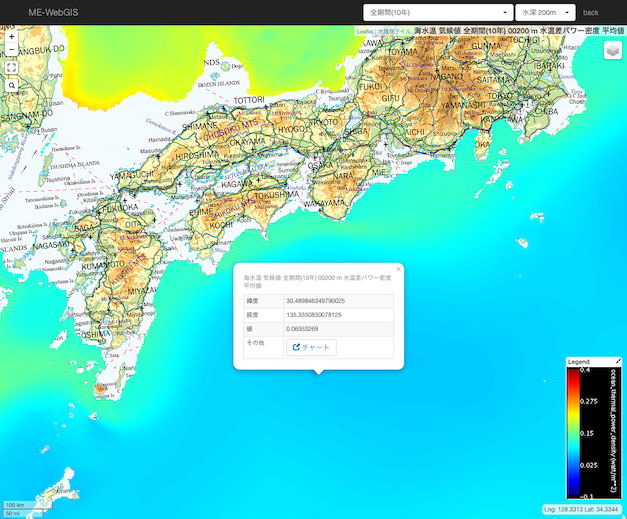}}
%
\caption{
Sample of JCOPE-T climatology map data available, covering up to a 10-year period (2002-2011).
%
%
Screenshots of interactive climatologies are shown in (a) and (b) at \SI{1}{\meter} depths for $P_{c}$ (10-year-seasonal) and $\bm u$ (large-meander) respectively, and (c) at a \SI{200}{\meter} depth for $P_{T}$ (10-year-annual).
%
%
}
\label{fig:OceanMap}
\end{figure}
%

The ocean and tidal current and ocean temperature resource assessments are accessible within the subcategory `Current and Temperature Map Information' (\url{http://www.todaiww3.k.u-tokyo.ac.jp/nedo_p/jp/webgis/#i-3}).
%
Six and three variables from each resource assessment (respectively) are chosen and analyzed for online distribution.
%
In each assessment, each statistic is calculated for nine depths, given by $\text{D}_c {=} [1, 50, 100, 200, 400, 500, 700, 1000]$ and $\text{D}_T {=} [5, 50, 100, 200, 400, 500, 700, 1000]$ (respectively). 
See Table~\ref{tab:OceanList} for a full list of variables.
%

For each variable and depth in Table~\ref{tab:OceanList}, climatologies are calculated for four different periods -- \emph{year}, \emph{season}, \emph{LM}, and \emph{NLM}.
As previously, `year' and `season' refer to annual and seasonal climatologies and cover the 10-year simulation. 
%
%
`LM' is shorthand for the large-meander Kuroshio path and covers the period 2004/07/01 to 2005/07/31; `NLM' is shorthand for the non-large-meander Kuroshio path and covers the simulation period 2002/01/01 to 2011/12/31 which excludes LM.

\begin{figure}[!t]
\centering
\subfigure[Spring depth profiles of $P_{c}$]
{\includegraphics[width=0.88\columnwidth,frame]{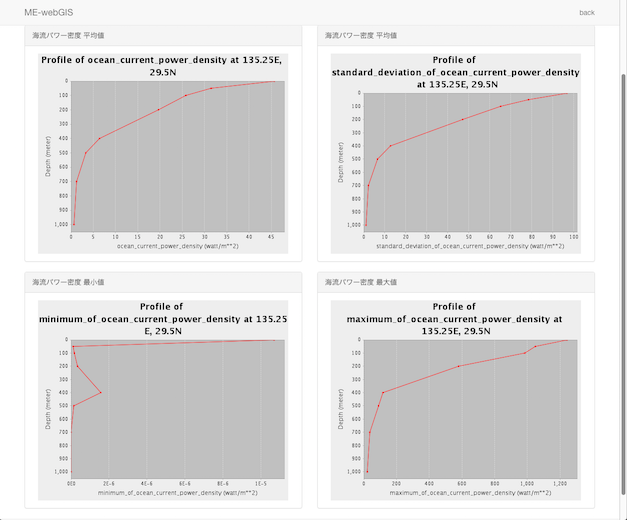}}
\hspace{0.05\columnwidth}
\subfigure[Spring frequency distribution of $\Delta T_{20}$]
{\includegraphics[width=0.88\columnwidth,frame]{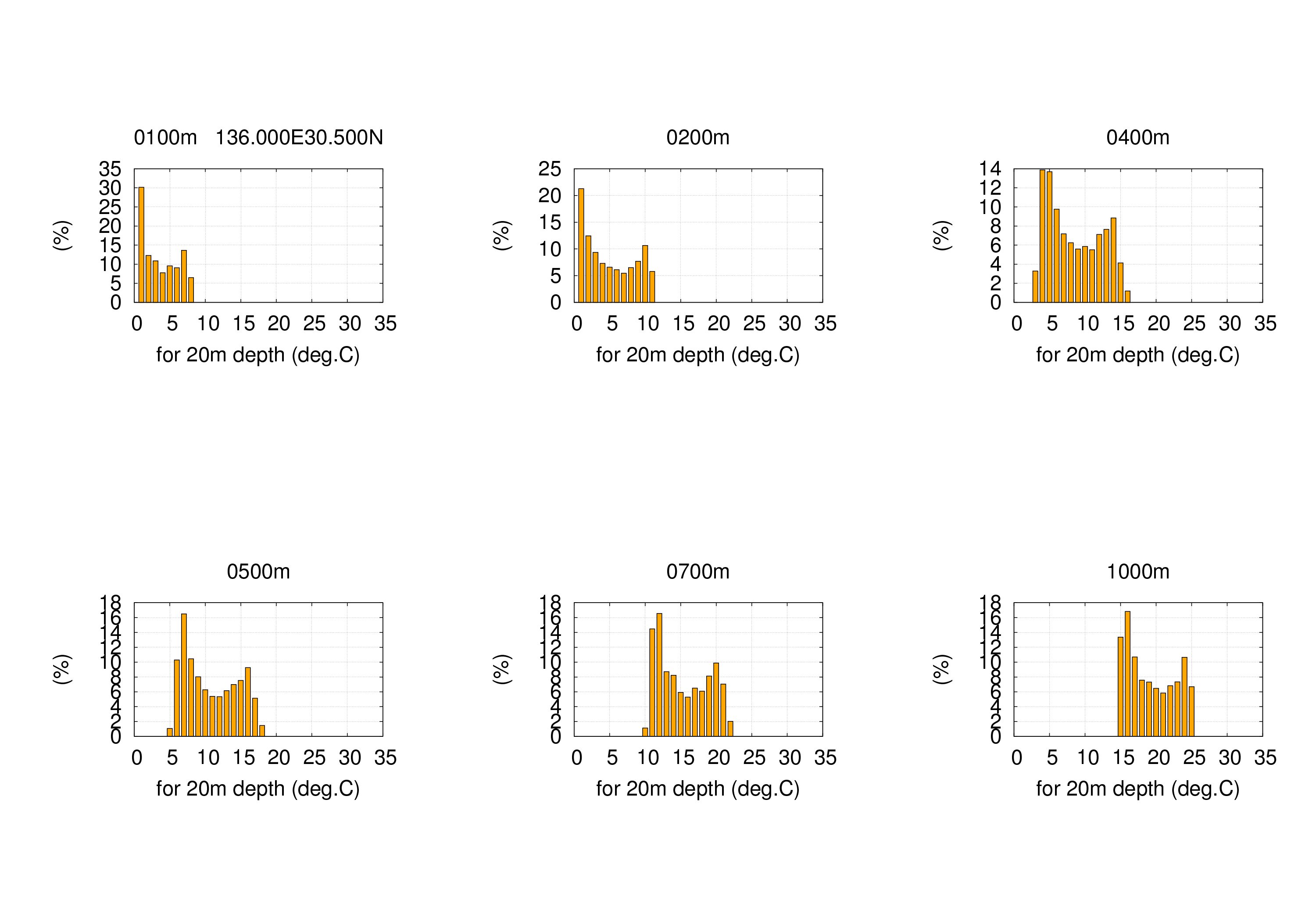}}
\caption{
Sample of JCOPE-T climatology profile and histogram data available at map locations for a 10-year period (2002--2011).
Screenshot of depth profiles is shown in (a) for seasonal $P_{c}$. 
%
Sample file download is shown in (b) of the seasonal $\Delta T_{20}$ frequency distribution histograms at specific depths.
%
%
%
}
\label{fig:OceanProfile}
\end{figure}

The four different ocean current and temperature climatologies are accessible as interactive 2D maps and depth profiles (at map locations); examples are shown in Figs.~\ref{fig:OceanMap}a--c and Fig.~\ref{fig:OceanProfile}a.
In addition, standard deviations, maximum and minimum values (within the different periods), and relative cumulative frequencies (greater than or equal) are available for select variables; see Table~\ref{tab:OceanProfile} for a list and the values used for $F_c$.
And finally, frequency distribution histograms are also available for select variables and can be downloaded as `png' and `csv' files. 
See also Table~\ref{tab:OceanProfile} for a list and Fig.~\ref{fig:OceanProfile}b for a sample download.

\begin{table}[!t]
\renewcommand{\arraystretch}{1.2}
\caption{
2D maps and depth profiles of JCOPE-T data are calculated for select period types and depths.
Different climatology types (year/season/LM/NLM) cover up to a 10-year period (2002--2011); `LM' and `NLM' are shorthand for the large-meander and non-large-meander Kuroshio paths.
Selected depths are represented by `D$_c$' and `D$_T$'.
Climatology of $\bm u$ refers to $\left( \langle u \rangle, \langle v \rangle \right)$.
See Section~\ref{sec:OceanMap} for details.
}
\label{tab:OceanList}
\centering\small
\begin{tabular}{p{0.1\columnwidth}
p{0.02\columnwidth}p{0.02\columnwidth}p{0.02\columnwidth}p{0.02\columnwidth}
p{0.02\columnwidth}p{0.02\columnwidth}p{0.02\columnwidth}p{0.02\columnwidth}
}
\toprule
\bf Variable 
& \multicolumn{6}{c}{\bf Climatology}
\\
%
%
& \multicolumn{4}{c}{\textit{Type}}
& \multicolumn{2}{c}{\textit{Depth}} 
\\
%
& \multicolumn{1}{c}{Year} & \multicolumn{1}{c}{Season} & \multicolumn{1}{c}{LM} & \multicolumn{1}{c}{NLM} 
& D$_c$ & D$_T$
\\
\midrule
$\lvert \bm u \rvert$
& \multicolumn{1}{c}{$\circ$} & \multicolumn{1}{c}{$\circ$} & \multicolumn{1}{c}{$\circ$} & \multicolumn{1}{c}{$\circ$} 
& \multicolumn{1}{c}{$\circ$} & 
\\
$\bm u$
& \multicolumn{1}{c}{$\circ$} & \multicolumn{1}{c}{$\circ$} & \multicolumn{1}{c}{$\circ$} & \multicolumn{1}{c}{$\circ$} 
& \multicolumn{1}{c}{$\circ$} & 
\\
$u$
& \multicolumn{1}{c}{$\circ$} & \multicolumn{1}{c}{$\circ$} & \multicolumn{1}{c}{$\circ$} & \multicolumn{1}{c}{$\circ$} 
& \multicolumn{1}{c}{$\circ$} & 
\\
$v$
& \multicolumn{1}{c}{$\circ$} & \multicolumn{1}{c}{$\circ$} & \multicolumn{1}{c}{$\circ$} & \multicolumn{1}{c}{$\circ$} 
& \multicolumn{1}{c}{$\circ$} & 
\\
$m_c$
& \multicolumn{1}{c}{$\circ$} & \multicolumn{1}{c}{$\circ$} & \multicolumn{1}{c}{$\circ$} & \multicolumn{1}{c}{$\circ$} 
& \multicolumn{1}{c}{$\circ$} & 
\\
%
$P_c$
& \multicolumn{1}{c}{$\circ$} & \multicolumn{1}{c}{$\circ$} & \multicolumn{1}{c}{$\circ$} & \multicolumn{1}{c}{$\circ$} 
& \multicolumn{1}{c}{$\circ$} & 
\\
\midrule
$T$
& \multicolumn{1}{c}{$\circ$} & \multicolumn{1}{c}{$\circ$} & \multicolumn{1}{c}{$\circ$} & \multicolumn{1}{c}{$\circ$} 
& & \multicolumn{1}{c}{$\circ$}  
\\
$\Delta T_{20}$
& \multicolumn{1}{c}{$\circ$} & \multicolumn{1}{c}{$\circ$} & \multicolumn{1}{c}{$\circ$} & \multicolumn{1}{c}{$\circ$} 
& & \multicolumn{1}{c}{$\circ$}  
\\
$P_T$
& \multicolumn{1}{c}{$\circ$} & \multicolumn{1}{c}{$\circ$} & \multicolumn{1}{c}{$\circ$} & \multicolumn{1}{c}{$\circ$} 
& & \multicolumn{1}{c}{$\circ$} 
\\
\bottomrule
\end{tabular}
\end{table}
\begin{table}[!t]
\renewcommand{\arraystretch}{1.2}
\caption{
List of 2D maps, depth profiles, and histograms available for JCOPE-T data.
`Clim' and `std' are shorthand for climatology and standard deviation;
`Max' and `Min' refer to the maximum and minimum values within the different periods. 
%
%
Values used for $F_c$ are listed in the table and frequency distribution histograms are downloadable.
See Section~\ref{sec:OceanMap} and Table~\ref{tab:OceanList} for further details.
%
%
%
%
}
\label{tab:OceanProfile}
\centering\small
\begin{tabular}{p{0.1\columnwidth}
p{0.02\columnwidth}p{0.02\columnwidth}p{0.02\columnwidth}p{0.02\columnwidth}p{0.02\columnwidth}
p{0.02\columnwidth}
}
\toprule
\bf Variable 
& \multicolumn{5}{c}{\bf Climatology} 
& \multicolumn{1}{c}{\bf Frequency} 
\\
& \multicolumn{5}{c}{\textit{Map / Profile}} 
& \multicolumn{1}{c}{\textit{Histogram}} 
\\
& \multicolumn{1}{c}{Clim} & \multicolumn{1}{c}{Std} & \multicolumn{1}{c}{Max} & \multicolumn{1}{c}{Min}
& \multicolumn{1}{c}{$F_c$} & \multicolumn{1}{c}{$F$} 
\\
\midrule
$\lvert \bm u \rvert$
& \multicolumn{1}{c}{$\circ$} & \multicolumn{1}{c}{$\circ$} & \multicolumn{1}{c}{$\circ$} & \multicolumn{1}{c}{$\circ$} 
& 
\multicolumn{1}{c}{$1$, $1.5$}
&
\\
$\bm u$
& \multicolumn{1}{c}{$\circ$} & & & &
&
\\
$u$
& \multicolumn{1}{c}{$\circ$} & & \multicolumn{1}{c}{$\circ$} & \multicolumn{1}{c}{$\circ$} &
&
\\
$v$
& \multicolumn{1}{c}{$\circ$} & & \multicolumn{1}{c}{$\circ$} & \multicolumn{1}{c}{$\circ$} &
&
\\
$m_c$
& \multicolumn{1}{c}{$\circ$} & & & &
&
\\
$\theta_c$
& & & & & & \multicolumn{1}{c}{$\circ$}
\\
$P_c$
& \multicolumn{1}{c}{$\circ$} & \multicolumn{1}{c}{$\circ$} & \multicolumn{1}{c}{$\circ$} & \multicolumn{1}{c}{$\circ$} &
&
\\
\midrule
$T$
& \multicolumn{1}{c}{$\circ$} & \multicolumn{1}{c}{$\circ$} & \multicolumn{1}{c}{$\circ$} & \multicolumn{1}{c}{$\circ$} &
& \multicolumn{1}{c}{$\circ$}
\\
$\Delta T_{\text{20}}$
& \multicolumn{1}{c}{$\circ$} & \multicolumn{1}{c}{$\circ$} & \multicolumn{1}{c}{$\circ$} & \multicolumn{1}{c}{$\circ$} & 
\multicolumn{1}{c}{$20$}
& \multicolumn{1}{c}{$\circ$}
\\
$P_T$
& \multicolumn{1}{c}{$\circ$} & \multicolumn{1}{c}{$\circ$} & \multicolumn{1}{c}{$\circ$} & \multicolumn{1}{c}{$\circ$} &
&
\\
\bottomrule
\end{tabular}
\end{table}
\section{Summary}
Marine renewable resource assessments conducted by The University of Tokyo and JAMSTEC are now publicly-available online at \url{http://www.todaiww3.k.u-tokyo.ac.jp/nedo_p/en}.
%
The data is distributed by a web GIS service that utilizes TDS and GeoServer software with Leaflet libraries.
%
The web GIS dataset contains statistical analyses of wave power (21 years), ocean and tidal current power (10 years), and ocean temperature power (10 years) that are accessed through interactive browser sessions and downloadable files.

\section*{Acknowledgment}
%
This work was supported by the NEDO project titled, 
``Research on the Framework and Infrastructure of Marine Renewable Energy; an Energy Potential Assessment.''
%
All sample screenshots utilize Geographical Survey Institute (GSI) English tile maps (\url{http://www.gsi.go.jp/}).





\bibliographystyle{IEEEtran}
%
%

\end{document}